\documentclass[prl,twocolumn,superscriptaddress,showpacs]{revtex4}
\usepackage{graphicx}
\begin{document}
\title{Crystal field and related phenomena in hexagonal close-packed H$_2$ and D$_2$ under pressure}
\author{Yu. A. Freiman}
\email{yuri.afreiman@gmail.com} \affiliation{B.Verkin Institute
for Low Temperature Physics and Engineering of the National
Academy of Sciences of Ukraine, 47 Lenin avenue, Kharkov, 61103,
Ukraine}
\author{S. M. Tretyak}
\affiliation{B.Verkin Institute for Low Temperature Physics and
Engineering of the National Academy of Sciences of Ukraine, 47
Lenin avenue, Kharkov, 61103, Ukraine}
\author{Alexei Grechnev}
\affiliation{B.Verkin Institute for Low Temperature Physics and
Engineering of the National Academy of Sciences of Ukraine, 47
Lenin avenue, Kharkov, 61103, Ukraine}
\author{Alexander F. Goncharov}
\affiliation{Geophysical Laboratory, Carnegie Institution of
Washington, 5251 Broad Branch Road NW, Washington, DC 20015, USA}
\affiliation{Center for Energy Matter in Extreme Environments and
Key Laboratory of Materials Physics, Institute of Solid State
Physics, Chinese Academy of Sciences, 350 Shushanghu Road, Hefei,
Anhui 230031, China}
\author{Russell J. Hemley}
\affiliation{Geophysical Laboratory, Carnegie Institution of
Washington, 5251 Broad Branch Road NW, Washington, DC 20015, USA}
\date{\today}

\begin{abstract}
The lattice distortion parameter (the deviation of the $c/a$ ratio
from the ideal value 1.633), orientational order parameter, and
crystal-field parameter in hexagonal close-packed (hcp) lattice of
$p$-H$_2$, $o$-D$_2$ and $n$-H$_2$ are calculated using the
semi-empirical lattice-dynamic approach. It is shown that the
lattice distortion in the $J$-even species is two order of
magnitude smaller compared with that found in $n$-H$_2$, and
$n$-D$_2$. The difference is due to the splitting of the $J$-odd
rotational levels in the $J$-even - $J$-odd mixtures.

\end{abstract}
\maketitle
\begin{large}

At zero pressure and temperature the molecules in $J$-even
($p$-H$_2$, $o$-D$_2$) and $J$-all (HD) solid hydrogens are in the
ground state $J=0$. Admixtures of higher rotational states into
the ground-state wave function are very small and the molecules
are virtually spherical. Rigid spheres crystallize into the face
center cubic (fcc) or hexagonal close-packed (hcp) lattices. As
compared with fcc,  hcp lattice has an additional degree of
freedom associated with the axial $c/a$ ratio. The lattice of
close-packed hard spheres has $c/a = \sqrt{8/3} \approx 1.633$ (an
ideal hcp structure). The quantity $\delta = c/a -\sqrt{8/3}$, the
lattice distortion parameter, describes the deviation of the axial
ratio from the ideal value. In the case of $\delta\, <\,0$, this
distortion involves an expansion within close-packed planes, and
contraction along the $c$-axis direction, and vice versa, for
$\delta\,>\, 0$ the lattice is expanded along the $c$-axis and
contracted within the close-packed planes.

Calculations with simple isotropic pair potentials have shown that
the ideal hcp lattice  at zero pressure and temperature does not
minimize the lattice energy
\cite{Howard70,Stillinger01,Schwerdtfeger06}, and the hcp lattice
with the minimal energy has a small but nonzero lattice
distortion. Semi-empirical and DFT calculations performed for
solid He  and hcp rare-gas solids (RGS) showed
\cite{Freiman09,Grechnev10} that the  pressure dependencies of the
lattice distortion parameter $\delta$  for the many-body (two-
plus three-body) and for the pair intermolecular potentials are
qualitatively different. The three-body forces flatten the lattice
($\delta<0$) while the pair forces at large compressions tend to
elongate it ($\delta>0$). Thus it was shown that
the lattice distortion parameter is a thermodynamic quantity,
which is very sensitive to the many-body component of the
intermolecular potential and can therefore be used as a probe of
the many-body forces \cite{Freiman02}.

The deviation of the axial ratio from the ideal value can be
attributed ultimately to a lowering of the band-structure energy
through lattice distortion. In the case of solid hydrogens the
effect of lattice distortion both on the isotropic and rotational
components of the ground-state energy is essential.

For all hcp elemental solids except helium, hydrogen, and
high-pressure Ar, Kr, and Xe the behavior of $\delta$ with
pressure and temperature is well established from both theory and
experiment. Typical values are of the order of 10$^{-2}$
\cite{Ashcroft76}. For solid helium $\delta$ is one-two orders of
magnitude smaller \cite{Freiman09,Grechnev10}.  The first
measurements of $c/a$ ratio in solid hydrogens  were done by
Keesom {\it et al.} \cite{Keesom30}, who found that at zero
pressure  the hcp lattice of $p$-H$_2$ is close to the ideal one.
X-ray zero-pressure study by Krupskii {\it et al.}
\cite{Krupskii83} confirmed this result ($c/a = 1.633\pm0.001$)
and extended it to the $o$-D$_2$. In fact, the only structural
study of $p$-H$_2$ and $o$-D$_2$ at elevated pressures up to 2.5
GPa and low temperatures were made by Ishmaev {\it et al.} using
the neutron diffraction method \cite{Ishmaev83,Ishmaev85}. It was
found that the ratio $c/a$ is practically constant and is slightly
lower than the ideal hcp value ($1.631 \pm 0.002$).

There were numerous structural studies of $n$-H$_2$ and $n$-D$_2$
\cite{Hazen87,Glazkov88,Mao88,Hu,Hemley90,Besedin90,Besedin91}
(see also reviews \cite{Mao94,Manzhelii97}). Using new synchrotron
x-ray diffraction techniques Mao {\it et al.} \cite{Mao88} and Hu
{\it et al.} \cite{Hu} found that in the case of $n$-H$_2$ the
pressure dependence of $c/a$ between 5 and 42 GPa is linear
\begin{equation}\label{c/a}
c/a=1.63-0.000441P,
\end{equation}
where $P$ is the pressure in GPa. Synchrotron single-crystal x-ray
diffraction measurements of $n$-H$_2$
\cite{Loubeyre96,Kawamura02,Akahama10} and $n$-D$_2$
\cite{Loubeyre96} up to megabar pressures at room temperature
revealed that the $c/a$ ratio decreases with increasing pressure
near linearly up to 180 GPa. No isotope effect in the pressure
dependence of the $c/a$ ratio was found.

There have been many attempts to calculate the pressure behavior
of the axial ratio using different theoretical approaches:
Hartree-Fock \cite{Anisimov77,Raynor87}, local density
approximation \cite{Barbee89}, path-integral Monte-Carlo
\cite{Cui97}, {\it ab initio} molecular dynamics
\cite{Kohanoff97}. In all proposed theoretical approaches the
rotation-lattice coupling gives rise to positive $\delta$ which is
in contradiction with experiment.

In the present paper we develop a self-consistent approach based
on the many-body semi-empirical intermolecular potential proposed
originally in Ref. \cite{Freiman01}. We are going to show that the
pressure behavior of the $c/a$-ratio in solid $p$-H$_2$ and
$o$-D$_2$ in phase I is determined mainly by the translational
degrees of freedom and is comparable in magnitude with that for
rare-gas solids. In the case of $n-$H$_2$ and $n-$D$_2$ we show
that the behavior of the axial ratio found in Refs.
\cite{Loubeyre96,Kawamura02,Akahama10} is due to the presence of
the $J$-odd component in normal ortho-para mixtures of solid
hydrogens \cite{Igarashi90}.

The Hamiltonian of the problem can be written in the form
\cite{Manzhelii97,Freiman01}
\begin{equation}\label{Hamiltonian}
{\cal H} = {\cal H}_{\rm is} + {\cal H}_{\rm rot} + {\cal H}_{\rm
int}\:,
\end{equation}
where ${\cal H}_{\rm is}$ is the contribution of the isotropic
part of the intermolecular potential, ${\cal H}_{\rm rot}$ is the
rotational part of the Hamiltonian, and ${\cal H}_{\rm int}$
describes the lattice--rotation coupling.

The isotropic part of the ground-state energy $E_{\rm is}$ can be
written as a sum of contributions from the isotropic part of the
static two-- and many--body energies  $E_{\rm is}^{\rm pair}$ and
$E_{\rm is}^{\rm m.b.}$ , and the zero--point vibrations
$E^{\mathrm{\rm zpv}}$:
\begin{equation}\label{(is)}
E_{\rm is} = E_{\rm is}^{\rm pair} + E_{\mathrm{is}}^{\rm m.b.} +
E^{\rm zpv}.
\end{equation}

In the mean field approximation (MFA), the Hamiltonian of the
system  of quantum linear rotors has the form \cite{Freiman01,
Manzhelii97}:
\begin{equation}\label{Hrot}
{\mathcal H}_{\rm rot} =  \sum_f{\bf L}_f^2 -
U_0\sqrt{4\pi/5}\,\eta\sum_fY_{20}(\vartheta_f) + \frac{1}{2}N
U_0\eta^2,
\end{equation}
where ${\bf L}_f$ is the angular momentum operator,
$Y_{LM}(\vartheta_f,\varphi_f)$ are spherical harmonics, and the
angles $\vartheta_f$ and $\varphi_f$ specify the orientational
axis of the molecule at the lattice site $f$. All  the energy
quantities are expressed in units of the rotational constant
$B_{\rm rot}$.

In distinction to the phases II and III, the phase I has no
orientational structure which would originate from the coupling
term in the anisotropic interaction between the hydrogen
molecules. A certain degree of orientational order in the phase I
as will be shown below originates from the crystal field
interaction. The orientational order parameter is defined as
\begin{equation}\label{eta}
\eta=\sqrt{4\pi/5}\langle{Y_{20}(\vartheta_f,\varphi_f)}\rangle,
\end{equation}
where $\langle\dots\rangle$ means averaging with the Hamiltonian
${\mathcal H}_{\rm rot}$ (Eq. \ref{Hrot}), and $N$ is the number
of sites.

The molecular field constant  is defined by
\begin{equation}\label{U0}
U_0=\sum_{ff'}\sum_{\alpha\beta\gamma\delta}V_{ff'}^{\alpha\beta\gamma\delta}
Q_f^{\alpha\beta}Q_{f'}^{\gamma\delta},
\end{equation}
where $V_{ff'}^{\alpha\beta\gamma\delta}$ is the interaction
matrix defined by the parameters of the intermolecular potential,
$Q_f^{\alpha\beta}=\Omega_f^{\alpha}\Omega_f^{\beta}-\frac{1}{3}\delta_{\alpha
\beta}$, ${\bf \Omega}$ is the unit vector specifying the
equilibrium orientation of the molecule at the site $f$. There is
a near linear correspondence between dimensionless pressure  in
units of $U_0/B_{\rm rot}$ and pressure in GPa. For the $Pca2_1$
lattice we have the following approximate relations for rescaling
the pressure: $P$ in $U_0/B_{\rm rot}$ units corresponds to 0.5$P$
in GPa for  H$_2$; 0.75$P$ for HD, and 1.25 for D$_2$
\cite{Freiman01}

The many-body hydrogen intermolecular potential  used here is a
sum of the pair SG potential \cite{Silvera78} (discarding the
$R^{-9}$ term) and two three-body terms: the long-range
Axilrod-Teller dispersive interaction and the short-range
three-body exchange interaction in the Slater-Kirkwood form
\cite{Loubeyre87,Freiman08}. It was successfully used for the
description of the equation of state, the pressure dependence of
the  Raman-active $E_{2g}$ mode \cite{Freiman12}, and the sound
velocities in solid hydrogen under pressure \cite{Freiman13}. The
explicit form and parameters of the potential used in this  work
are given in Ref. \onlinecite{Freiman11}. The contribution of the
zero--point vibrations $E^{\mathrm{\rm zpv}}$ was taken into
account in the Einstein approximation.

In a non-rigid lattice there is a strong lattice-rotation
coupling, which is due to the crystal field. The origin of this
coupling can be explained in the following way. With increasing
pressure the anisotropic interaction increases and admixtures of
$J \neq 0$ rotational states  into the ground state wave function
become more and more appreciable. With the nonzero admixture the
molecules acquire anisotropy. The anisotropic molecules tend to be
packed into a distorted lattice. The lattice distortion $\delta$
is given by a competition of the anisotropic interactions (which
favor strong distortion) and the isotropic interactions (which
favor a near-ideal HCP lattice).

The lattice-rotation coupling is described by the the Hamiltonian
\begin{equation}\label{Hint}
{\cal H}_{\rm int}= -\varepsilon_{2c}\sqrt{4\pi/5}Y_{20},
\end{equation}
where $\varepsilon_{2c}$ is the crystal-field parameter
\cite{Kranendonk83} which is linear with respect to $\delta$:
\begin{equation}\label{epsilon}
\varepsilon_{2c}=\tilde{B}\delta;\,\,\,\,\,
\tilde{B}=-\sqrt{6}\left(B+\frac{1}{2}R\frac{dB}{dR}\right),
    \end{equation}
where $B(R)$ is the radial function  of the single-molecular term
in the anisotropic intermolecular potential \cite{Kranendonk83}.
Thus, the state of the lattice can be described by two coupled
order parameters, $\eta(V, T)$ and $\delta(V, T)$, which can be
determined by the minimization of the free energy with respect to
these parameters. In the calculations we restrict ourselves  to
$T=0$ K case, so we will minimize the total ground-state energy
$E_0^{\rm tot}=E_0^{\rm tr}+ E_0^{\rm rot}$ (the superscripts
"tot", "tr", and "rot" refer to the total ground-state energy, and
translational, and rotational subsystems, respectively))

The translational part of the round-state energy $E_0^{\rm tr}$
does not depend on $\eta$ and respective minimum conditions take
the form:
\begin{equation}\label{minimum condition 1}
\partial\, E_0^{\rm rot}/\partial\,\eta = 0;;\,\,\,\,\,
    \end{equation}

\begin{equation}\label{minimum condition 2}
\partial\,\left(E_0^{\rm tr}+ E_0^{\rm rot}\right)/\partial\, \delta =
0.
\end{equation}

Thus, the complete minimization can be carried out in two stages,
first, with the help of Eq. (\ref{minimum condition 1}) we find
$\eta$ as a function of $U_0$ and $\delta$ and then by minimizing
the total ground-state energy with respect to $\delta$ (Eq.
\ref{minimum condition 2}) we find $\delta$ and $\eta$ as a
function of $U_0$ (volume $V$).

Using the successive approximation method we can find solutions of
Eqs. (\ref{minimum condition 1}), (\ref{minimum condition 2}) in
any necessary approximation \cite{Freiman01,Freiman11}. Up to the
third order in the crystal-field parameter  the orientational
order parameter and orientational ground-state energy have the
following form:
\begin{equation}
\label{eta2}
\eta=\kappa\,\frac{\epsilon_{2c}}{B_{\rm rot}}+
\frac{15^2}{14}\kappa^3\left(\frac{\epsilon_{2c}}{B_{\rm
rot}}\right)^2.
\end{equation}
\begin{equation}\label{epsilon1}
 \frac{E_0^{\rm rot}}{B_{\rm rot}}= -\frac{1}{2}\kappa\,\left(\frac{\epsilon_{2c}}{B_{\rm rot}}\right)^2
-\frac{75}{14}\kappa^3\left(\frac{\epsilon_{2c}}{B_{\rm
rot}}\right)^3.
\end{equation}
where
\begin{equation}\label{kappa}
 \kappa = \frac{1}{15-U_0/B_{\rm rot}}.
\end{equation}

The expansion parameter $\epsilon_{2c}$ (Eq. \ref{epsilon}) is
negative for all pressures, so the expansions Eqs. (\ref{eta2}),
(\ref{epsilon1}) are oscillating  and converge if the terms of the
expansions are decreasing.

Due to the presence of a singular factor $\kappa$ (Eq.
\ref{kappa}) in the expansions Eqs. (\ref{eta2}, \ref{epsilon1}),
the validity of this analytical solution is limited by the
condition $U_0/B_{\rm rot}<15$, which corresponds to pressure of
$\sim30$\,GPa for the case of $p$-H$_2$, and 20 GPa and 12 GPa for
HD and $o-$D$_2$, respectively. To extend the solution into the
higher pressure region a numerical approach should be used.

As can be seen from Eqs. (\ref{eta2}), (\ref{epsilon1}), at low
pressures the rotational part of the ground-state energy contains
no linear term in $\delta$. The same is also true at high
pressures. The admixture of higher rotational states to the
rotational wavefunction is proportional to the lattice distortion:
\begin{equation}\label{Psi}
\Psi_0 = Y_{00} + c_2\delta Y_{20}+...
\end{equation}

The ground-state energy
\begin{equation}\label{Erot}
E_0^{\rm rot}= \langle\Psi _0|{\mathcal H}_{\rm
rot}|\Psi_0\rangle=6c_2^2\delta^2-(U_0\eta^2/2+\epsilon_{2c}\eta).
\end{equation}
Since $\eta\sim\delta$ and $\epsilon_{2c}\sim\delta$, $E_0^{\rm
rot}\sim\delta^2$ at all pressures.

Let us consider the contribution of the translational degrees of
freedom. Up to terms of the second order in $\delta$ the
translational part of the ground-state energy is
\begin{equation}\label{translation}
 E_0^{\rm tr}=E_0(0) + b_1^{\rm tr}\,\delta + b_2^{\rm tr}\,\delta^2,
\end{equation}
where $E_0$ is the ground-state energy of the ideal lattice and
$b_i^{\rm tr}\,  (i=1, 2)$ are the coefficients which depend on
the parameters of the intermolecular potential and molar volume.
The total ground-state energy is
\begin{equation}\label{Etot}
E_0^{\rm tot}=  E_0(0) + b_1^{\rm tr}\,\delta + b_2^{\rm
tot}\,\delta^2,
\end{equation}
where $b_2^{\rm tot}= b_2^{\rm tr} + b_2^{\rm rot}.$
 Minimizing $E_0(\delta)$ over $\delta$, we
obtain
\begin{equation}\label{delta}
\delta = -\, b_1/(2\,b_2^{\rm tot}).
\end{equation}

\begin{figure}[!t]
 %%%   \centering
    {\includegraphics[width=9.0cm]{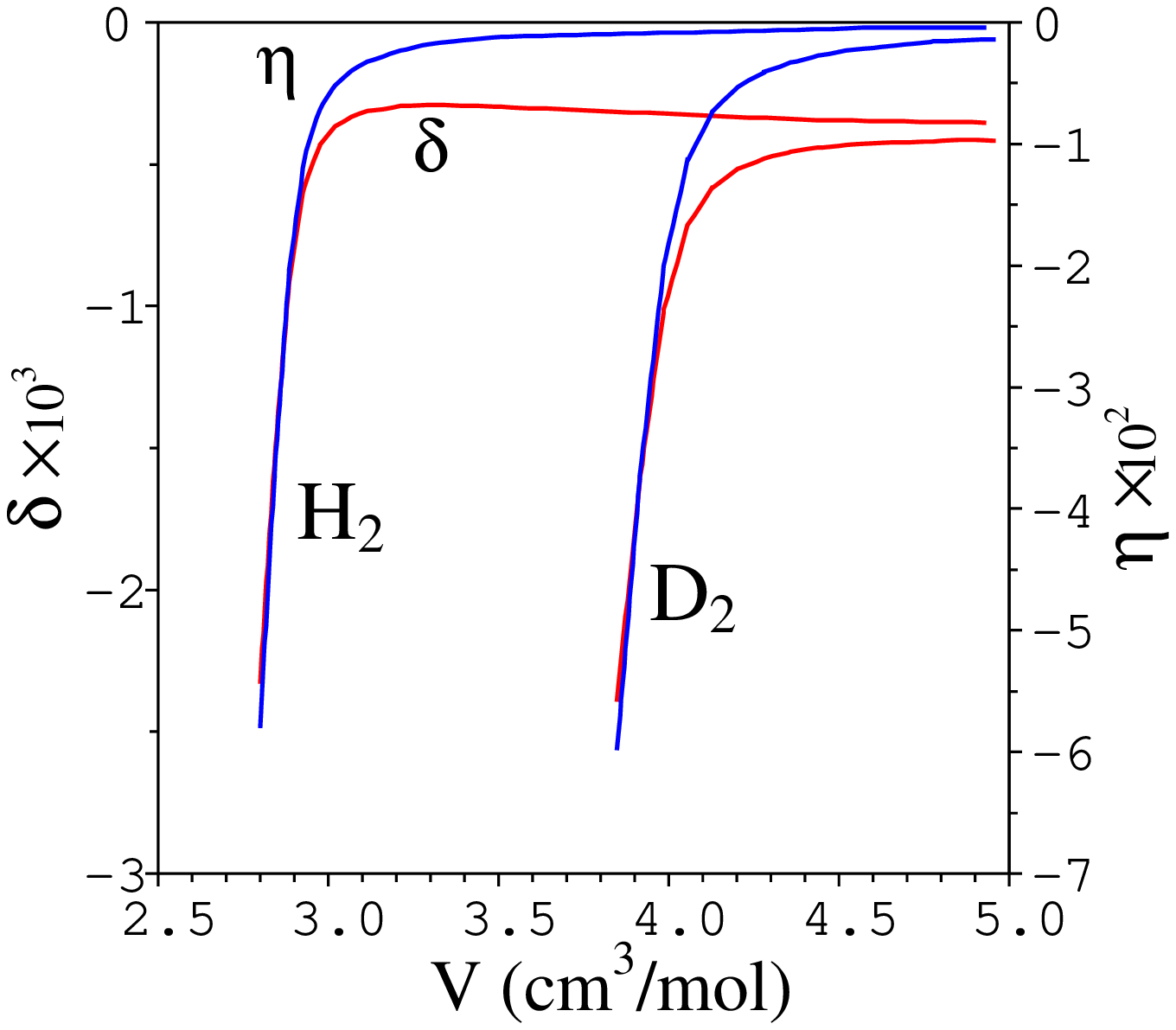}}
            \caption{\label{Fig1} \small{Lattice distortion
            parameter $\delta$ and orientational order parameter
            $\eta$ in  parahydrogen and orthodeuterium as
            functions of molar volume.}}
            {\includegraphics[width=9.0cm]{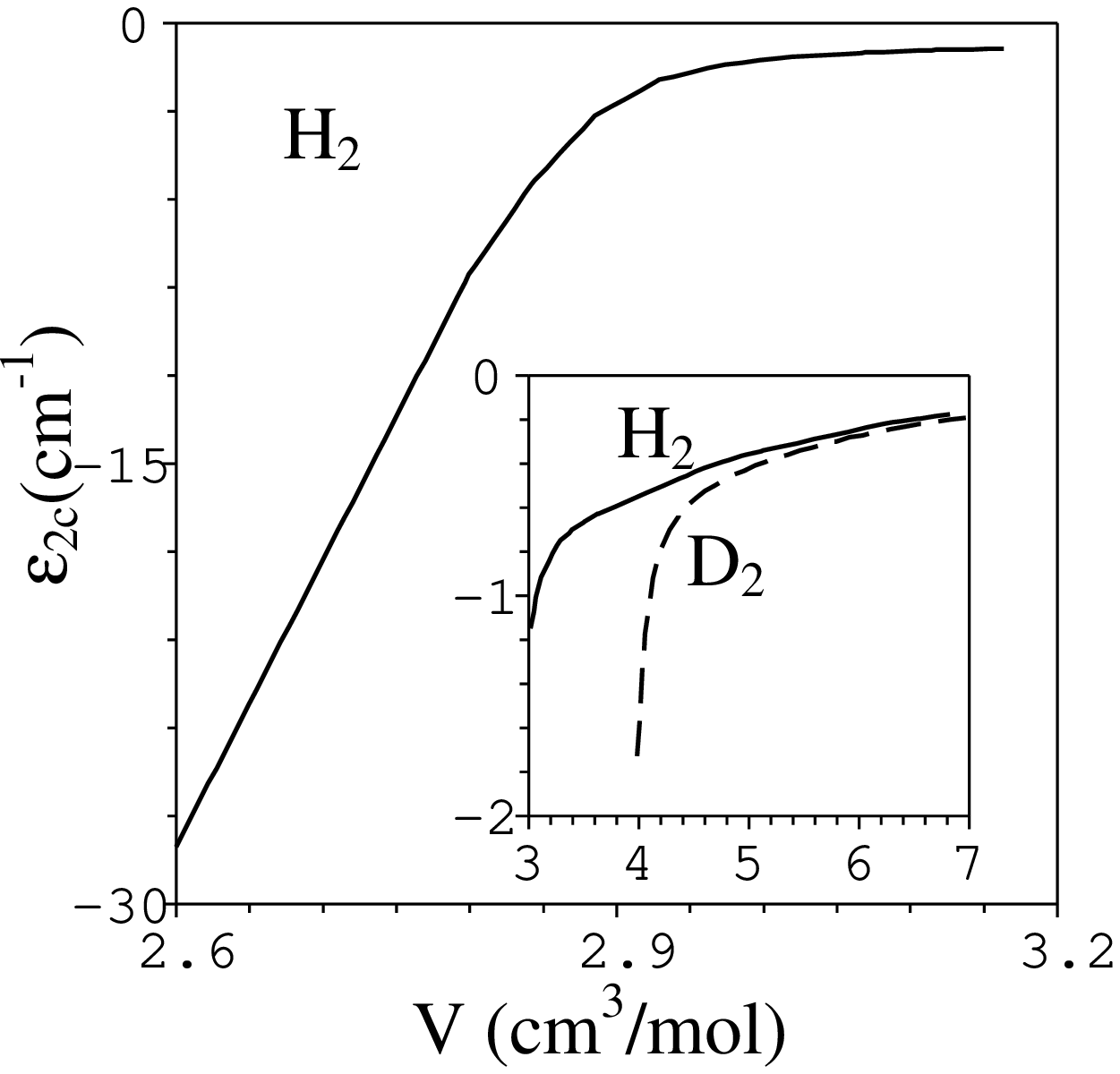}}
            \caption{\label{Fig2} \small{Crystal field parameter
            $\epsilon_{2c}$ in  parahydrogen and orthodeuterium (inset) as
            functions of molar volume.}}
            \end{figure}

At small pressures the contribution of the rotational degrees of
freedom to $b_2$ as follows from Eq. (\ref{epsilon1}) is $b_2^{\rm
rot}=-\kappa\left(\epsilon_{2c}^2/2B_{\rm rot}\right)$. It is
negative and increases in the absolute value with pressure. The
total $b_2^{\rm tot}$ is a sum of the respective contributions
\begin{equation}\label{b2rot}
b_2^{\rm tot} = b_2^{\rm tr} + b_2^{\rm rot}.
\end{equation}
Figure \ref{Fig1} presents the lattice distortion parameter
$\delta$ and orientational order parameter $\eta$ for $p$-H$_2$,
and $o-$D$_2$ as functions of molar volume. The calculations were
performed for the molar volumes $V<18\,\,\, {\rm cm}^3$/mol ($P>
0.5$ GPa) outside of the region where quantum crystal effects are
decisive and extended up to the point of the I - II transition. In
phase I both $\eta$ and $\delta$ are small and negative. The
negative $\delta$ means that the lattice is slightly flattened
compared with the ideal one; the negative $\eta$ means that the
molecules precess around the $c$-axis with the molecular axis
inclined to the $c$ axis by the angle slightly over
$\langle\vartheta_0\rangle=\cos^{-1}{1/\sqrt3}\approx
54^{\circ}44'$. Upon increasing pressure, $\eta$ decreases
monotonically (the limiting value $\eta=-1/2$ means that the
molecules precess around the $c$-axis with the precession angle
$\vartheta=\pi/2$). At large molar volumes (~18 cm$^3$/mol) the
deviation of the molecular ground state from the pure spherical
one is very small. This deviation is characterized by the
orientational order parameter $\eta=-3.3\cdot10^{-4}$ for
$p-$H$_2$ and  $\eta= -1\cdot10^{-5}$ for $o-$D$_2$. The lattice
is very close to the ideal one. While $\eta$ decreases
monotonically with rising pressure, $\delta$ changes with pressure
nonmonotonically (Figure \ref{Fig1}). This nonmonotonic behavior
is connected with the mutual changes of the coefficients $b_1^{\rm
tr},\,b_2^{\rm tr}$, and $b_2^{\rm rot}$ in Eq. (\ref{delta}) with
pressure.

Significant changes are seen between $V$ = 3.1 ($P$= 80 GPa) and
2.93 ($P$= 92.2 GPa) cm$^3$/mol for H$_2$ and between $V$= 4.05
(37.45 GPa) and 4.2 (33.8 GPa) for D$_2$, which corresponds to the
I-II transition (experimentally, 110 GPa in $p-$H$_2$
\cite{Lorenzana90}, 28 GPa GPa in $o-$D$_2$ \cite{Silvera81}).

Knowing the pressure dependence of $\delta$, we were able to
calculate the pressure dependence of the second order crystal
field parameter $\epsilon_{2c}$ (Fig. \ref{epsilon}). As known
\cite{Kranendonk83}, this parameter determines the splitting of
the purely rotational band $S_0(0)$ in $p-$H$_2$ and $o-$D$_2$ and
splitting of the rotational levels of impurity $J=1$ molecules in
$J=0$ solids. In the absence of direct experimental data some
qualitative conclusions on the lattice distortion parameter of
$p$-H$_2$ were obtained by Goncharov {\it et al.}
\cite{Goncharov01}. The authors have measured low-frequency Raman
spectra at low temperature for the pressure range up to the I - II
phase transition and used these spectra to estimate the
crystal-field parameter $\varepsilon_{2c}$. Assuming that only the
second-order crystal field is responsible for the splitting of the
roton triplet band $S_0(0)$ the authors obtained $\mid
\varepsilon_{2c}\mid \sim$ 1 cm$^{-1}$ and thus $\mid\delta \mid
\sim 10^{-3}-10^{-4}$ in accord with theoretical data shown in
Figs. (1, 2).

Characteristics of the rotational motion of the molecules in
$p-$H$_2$ and $o-$D$_2$  could be compared with the ones obtained
using the technique which combines a density functional theory
(DFT) with a path-integral molecular dynamics (PIMD)
\cite{Geneste12}. A direct comparison of the results involves
difficulties because the authors used the definition of the
orientational order parameter
($\eta=[N^{-1}\sum_i^N\sqrt{4\pi/5}Y_{20}({\bf \Omega}_i\cdot{\bf
u_i})]^2$ where ${\bf \Omega}_i$ is a unit vector specifying the
equilibrium orientation of the molecule at the site $i$, and ${\bf
u}_i$ is a site-specific unit vector which defines the
orientational structure),  which excludes negative values of the
order parameter. Nonetheless, the pressure evolutions of the order
parameters in the both approaches are similar.

Let us turn to the case of $n-$H$_2$. For a single $J = 1$
molecule in the lattice of $J=0$ molecules there is an additional
contribution to the ground-state energy arising from the
polarization of the surrounding $J=0$ molecules by the the
electric quadrupole field of the $J=1$ molecule. The polarization
energy due to the interaction of the quadrupole moment of the
$J=1$ molecule with the induced dipole moments of the surrounding
nearest neighboring $J=0$ molecules is equal to
\cite{Kranendonk83}
\begin{equation}\label{Eq5.3.2}
\epsilon_1=-18\alpha Q^2 V^{-8/3},
    \end{equation}
where $\alpha$ is the polarizability of the $J=0$ molecules and
$Q$ is the quadrupole moment of the $J=1$ molecule.

If the crystal shows a homogeneous deviation from the ideal hcp
structure specified by the lattice distortion parameter $\delta$,
the polarization energy contains a crystal-field term
\cite{Kranendonk83}
\begin{equation}\label{Eq5.3.3}
V_c=\varepsilon_{2c}\,\sqrt{\frac{4\pi}{5}}\,Y_{20}({\vec{\Omega}}),
    \end{equation}
where ${\vec{\Omega}}$ specifies the orientation of the $J=1$
molecule with
\begin{equation}\label{Eq5.3.4}
\varepsilon_{2c}= - \frac{24}{7}\epsilon_1\delta.
    \end{equation}

\begin{figure}[!t]
    \centering
        {\includegraphics[width=9.0cm]{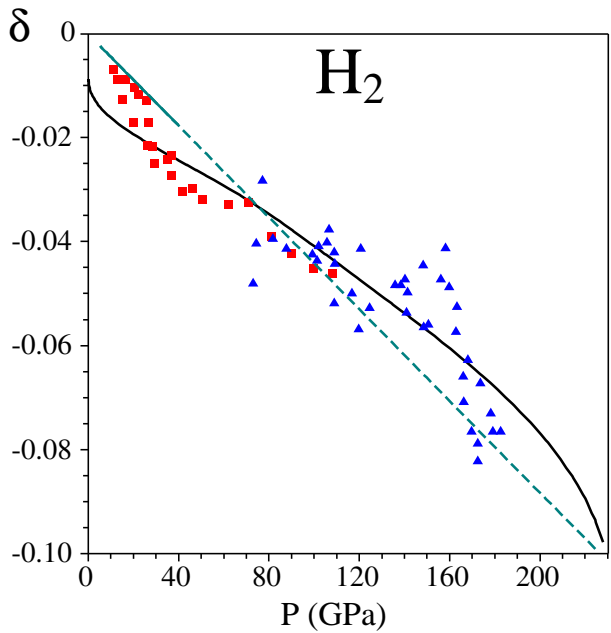}}
        \caption{\label{Fig3} \small{Lattice distortion parameter in normal
        ortho-para mixture of solid hydrogens. Theory: solid line
        - this work; experiment: red squares - data from Ref.
        \cite{Loubeyre96}, blue triangles - data from Ref.
        \cite{Akahama10},
         green straight line: Eq. (\ref{c/a}) (solid section \cite{Mao88,Hu},
        dashed section - extrapolation to high pressures.)}}
     \end{figure}

The triplet  $J=1$ level is splitted in the crystal field $V_c$;
the splitting is given by
\begin{equation}\label{Eq5.3.5}
\Delta_c= E(\pm) - E(0) = \frac{3}{5}\,|\epsilon_{2c}|,
    \end{equation}
where $E(M)$ is the energy of the state $J=1$, $J_z=M$ with $z$
direction parallel to the $c$ axis of the crystal. The positive
sign of $\Delta_c$ implies that the state $J_z=0$ is the ground
state of the triplet.  Due to this splitting the ground-state
energy is brought down by $2|\varepsilon_{2c}|/5$. The gain in the
anisotropic energy due to the polarization effect in the $J=0 /
J=1$ mixture is linear in $\delta$, contrary to the contribution
quadratic in delta in $J$-even solids. At the same time, this
distortion introduces an additional positive contribution to the
ground-state energy from the isotropic part of the intermolecular
interaction. This contribution is quadratic in the lattice
distortion parameter $\delta$ \cite{Freiman11}. The loss in the
isotropic part and the gain in the anisotropic part of the
ground-state energy determines the lattice distortion parameter at
the given molar volume. One cannot obtain a reliable value of
$\delta$ from Eqs. (\ref{Eq5.3.2}) - (\ref{Eq5.3.5}) because there
are other contributions to the splitting $\Delta_c$ not included
in Eq. (\ref{Eq5.3.5}) \cite{Kranendonk83}. Moreover, the lattice
of $n-$H$_2$ is a disordered mixture of ortho  and para molecules.
To calculate the polarization energy in such lattice we used a
mean-field model of ortho-para mixture assuming that each site of
the lattice is occupied by a superposition of even-$J$ and odd-$J$
molecule and the crystal field splitting occurs for every
ortho-component of such molecule. Assuming that the volume
dependence of the polarization energy has the same form as in Eq.
\ref{Eq5.3.2}) we included into the ground-state energy the term
\begin{equation}\label{E_pol}
E_{\rm pol}= A(V_0/V)^{8/3}\delta,
\end{equation}
where $A$ is an adjusting parameter of the theory. Comparing with
the experimental data on the pressure dependence of the lattice
distortion parameter from Refs. \cite{Loubeyre96,Akahama10} we
obtain a good fit for $A= 5.1$ K. The resulting pressure
dependence of the lattice distortion parameter is presented in
Fig. \ref{Fig3}. Comparing this result with the one for $p-$H$_2$
(Fig. \ref{Fig1}), we see that the introduction of the $J=1$
molecules increases the hcp lattice distortion by two orders of
magnitude.

In conclusion, we developed the lattice-dynamics theory of the
lattice distortion in $p-$hydrogen, $o-$deuterium, and
$n-$hydrogen in hexagonal closed-packed lattice under pressure. It
is shown that the lattice distortion in $p-$H$_2$ and $o-$D$_2$ is
negative and very small (of the order of 10$^{-3}$)
and their lattices are very close to the ideal one. In this aspect
the $J$-even modifications of hydrogens are very similar to solid
helium.  The main contributions to the lattice distortion comes
from translational degrees of freedom, although there is a nonzero
contribution from rotational degrees of freedom as well. The
negative sign of the lattice distortion parameter means that the
lattice is slightly flattened. The orientational order parameter
was calculated and shown to be small and negative. The molecules
rotate around the $c$-axis with the inclination angle $\vartheta
\approx55^{\circ}$.

A mean-field model of the ortho-para mixture of solid hydrogen was
developed to calculate the polarization energy connected with the
impurity ortho molecules. It is shown that there is a considerable
gain in the ground-state energy of the mixture due to the
polarization energy, which makes the lattice distortion
advantageous. The corresponding loss in the energy of the
isotropic interaction determines the resulting value of the
lattice distortion parameter. Obtained pressure dependence of the
$c/a$ ratio is in an excellent agreement with the experiment in
the whole pressure range.
\begin{flushleft}

\end{flushleft}
\end{large}
\end{document}